\documentclass[preprint2]{aastex631}
\usepackage{amsmath,amstext}
\usepackage[T1]{fontenc}
\usepackage[figure,figure*]{hypcap}
\usepackage{multirow}
\usepackage[T1]{fontenc} 
\usepackage[utf8]{inputenc} 

\newcommand{\UA}{Steward Observatory, The University of Arizona, 933 N.\ Cherry Ave, Tucson, AZ 85721, USA}
\newcommand{\PSUAA}{Department of Astronomy \& Astrophysics, The Pennsylvania State University, 525 Davey Laboratory, University Park, PA 16802, USA}
\newcommand{\PSUCEHW}{Center for Exoplanets and Habitable Worlds, The Pennsylvania State University, 525 Davey Laboratory, University Park, PA 16802, USA}
\newcommand{\Princeton}{Department of Astrophysical Sciences, Princeton University, 4 Ivy Lane, Princeton, NJ 08540, USA}
\newcommand{\Macquarie}{Department of Physics and Astronomy, Macquarie University, Balaclava Road, North Ryde, NSW 2109, Australia}
\newcommand{\UCI}{Department of Physics \& Astronomy, The University of California, Irvine, Irvine, CA 92697, USA}

\newcommand{\NIST}{Time and Frequency Division, National Institute of Standards and Technology, 325 Broadway, Boulder, CO 80305, USA}
\newcommand{\CUBoulder}{Department of Physics, University of Colorado, 2000 Colorado Avenue, Boulder, CO 80309, USA}
\newcommand{\UT}{Center for Planetary Systems Habitability and McDonald Observatory, The University of Texas at Austin, Austin, TX 78730, USA}

\newcommand{\JPL}{Jet Propulsion Laboratory, California Institute of Technology, 4800 Oak Grove Drive, Pasadena, California 91109, USA}
\newcommand{\SVD}{Space Vehicles Directorate, Air Force Research Laboratory, 3550 Aberdeen Ave SE, Kirtland AFB, NM 87117 USA}
\newcommand{\STSCI}{Space Telescope Science Institute, 3700 San Martin Drive, Baltimore, MD 21218, USA}
\newcommand{\JH}{Department of Physics and Astronomy, Johns Hopkins University, 3400 N Charles St, Baltimore, MD 21218, USA}

\newcommand{\unit}[1]{\ensuremath{\, \mathrm{#1}}}

\shortauthors{Ca\~nas et al.}
\shorttitle{A Mars-sized USP}

\begin{document}
\title{A hot Mars-sized exoplanet transiting an M dwarf}
\correspondingauthor{Caleb I. Ca\~nas}
\email{canas@psu.edu}

\author[0000-0003-4835-0619]{Caleb I. Ca\~nas}
\altaffiliation{NASA Earth and Space Science Fellow}
\affiliation{\PSUAA}
\affiliation{\PSUCEHW}

\author[0000-0001-9596-7983]{Suvrath Mahadevan}
\affil{\PSUAA}
\affil{\PSUCEHW}

\author[0000-0001-9662-3496]{William D. Cochran}
\affil{\UT}

\author[0000-0003-4384-7220]{Chad F.\ Bender}
\affil{\UA}

\author[0000-0002-5077-6734]{Eric D. Feigelson}
\affil{\PSUAA}
\affil{\PSUCEHW}

\author[0000-0003-2281-1990]{C. E. Harman}
\affil{NASA Ames Research Center, Moffett Field, CA 94035, USA}

\author[0000-0002-5893-2471]{Ravi Kumar Kopparapu}
\affiliation{Planetary Environments Laboratory, NASA Goddard Space Flight Center, Greenbelt, MD 20771, USA}

\author[0000-0002-2947-2023]{Gabriel A. Caceres}
\affil{Teachers Pay Teachers, 111 East 18th St, New York, NY 10003, USA}

\author[0000-0002-2144-0764]{Scott A. Diddams}
\affil{\NIST}
\affil{\CUBoulder}

\author[0000-0002-7714-6310]{Michael Endl}
\affil{\UT}

\author[0000-0001-6545-639X]{Eric B. Ford}
\affil{\PSUAA}
\affil{Institute for Computational \& Data Sciences, The Pennsylvania State University, University Park, PA 16802, USA}
\affil{\PSUCEHW}
\affil{Center for Astrostatistics, 525 Davey Lab, The Pennsylvania State University, University Park, PA 16802, USA}

\author[0000-0003-1312-9391]{Samuel Halverson}
\affil{\JPL}

\author[0000-0002-1664-3102]{Fred Hearty}
\affil{\PSUAA}
\affil{\PSUCEHW}

\author[0000-0002-7227-2334]{Sinclaire Jones}
\affiliation{\Princeton}

\author[0000-0001-8401-4300]{Shubham Kanodia}
\affiliation{\PSUAA}
\affiliation{\PSUCEHW}

\author[0000-0002-9082-6337]{Andrea S.J. Lin}
\affil{\PSUAA}
\affil{\PSUCEHW}

\author[0000-0001-5000-1018]{Andrew J. Metcalf}
\affil{\SVD}

\author[0000-0002-0048-2586]{Andrew Monson}
\affil{\PSUAA}

\author[0000-0001-8720-5612]{Joe P. Ninan}
\affiliation{\PSUAA}
\affiliation{\PSUCEHW}

\author[0000-0002-4289-7958]{Lawrence W. Ramsey}
\affil{\PSUAA}
\affil{\PSUCEHW}

\author[0000-0003-0149-9678]{Paul Robertson}
\affil{\UCI}

\author[0000-0001-8127-5775]{Arpita Roy}
\affil{\STSCI}
\affil{\JH}

\author[0000-0002-4046-987X]{Christian Schwab}
\affil{\Macquarie}

\author[0000-0001-7409-5688]{Gu\dh mundur Stef\'ansson}
\altaffiliation{Henry Norris Russell Fellow}
\affiliation{\Princeton}

\begin{abstract}
We validate the planetary nature of an ultra-short period planet orbiting the M dwarf KOI-4777. We use a combination of space-based photometry from Kepler, high-precision, near-infrared Doppler spectroscopy from the Habitable-zone Planet Finder, and adaptive optics imaging to characterize this system. KOI-4777.01 is a Mars-sized exoplanet ($\mathrm{R}_{p}=0.51 \pm 0.03R_{\oplus}$) orbiting the host star every \(0.412\)-days (\(\sim9.9\)-hours). This is the smallest validated ultra-short period planet known and we see no evidence for additional massive companions using our HPF RVs. We constrain the upper $3\sigma$ mass to $M_{p}<0.34~\mathrm{M_\oplus}$ by assuming the planet is less dense than iron. Obtaining a mass measurement for KOI-4777.01 is beyond current instrumental capabilities.

\end{abstract}


\section{Introduction}
Ultra-short period planets (USPs) have orbital periods \(<1\) day and represent a rare class of exoplanet. A statistical study of the Kepler data \citep{Sanchis-Ojeda2014} revealed 106 USP candidates and a dependence of the USP occurrence rate with the host star mass. \cite{Sanchis-Ojeda2014} calculated the occurrence rate of USPs to be \(0.15\pm0.05\%\) for F dwarfs with a maximum occurrence rate of \(1.10\pm0.40\%\) for M dwarfs. The Kepler \citep{Borucki2010} and K2 \citep{Howell2014} missions discovered only a few USPs transiting M dwarfs \citep[e.g.,][]{Muirhead2012,Swift2013,Sanchis-Ojeda2015,Smith2018,Hirano2018}. The first three years of the Transiting Exoplanet Survey Satellite \citep[TESS;][]{Ricker2015} have almost doubled the number of known USPs transiting M dwarfs with discoveries such as TOI-136 b \citep{Vanderspek2019}, TOI-732 b \citep{Cloutier2020,Nowak2020}, TOI-736 b \citep{Crossfield2019}, TOI-1078 b \citep{Shporer2020}, TOI-1634 b \citep{Cloutier2021,Hirano2021}, TOI-1635 b \citep{Hirano2021}, and TOI-1685 \citep{Bluhm2021}. 

The formation mechanisms for USPs are not completely known, but many proposed formation scenarios invoke inward migration because the location of observed USPs is interior to the dust sublimation radius \citep{Swift2013} and these regions would lack the materials necessary for \textit{in-situ} planet formation. Early formation scenarios proposed high-eccentricity migration in multi-planet systems \citep[e.g.,][]{Schlaufman2010} where dynamical interactions with planets on wider orbits would excite the eccentricity of progenitor USP until tidal interactions with the host star become strong enough to decay the orbit. Other proposed scenarios include (i) low-eccentricity migration due to secular planet-planet interactions \citep{Pu2019}, (ii) high-eccentricity migration due to chaotic secular interactions in compact multi-planet systems \citep{Petrovich2019}, (iii) tidal migration of USPs formed \textit{in situ} in truncated planetary disks \citep{Lee2017}, (iv) obliquity-driven tidal migration \citep{Millholland2020}, and (v) migration of the outer planet in a resonant chain towards the inner edge of a gaseous protoplanetary disc due to gravitational instability \citep{Zawadzki2021}. 

A majority of USPs are found in multi-planet systems \citep[e.g.,][]{Sanchis-Ojeda2014,Adams2017,Winn2018} and, when observed in multi-planet systems, USPs have larger period ratios with their nearest neighbor when compared to the period ratios between neighboring planets of the same system \citep{Steffen2013,Winn2018}, and larger mutual inclinations than when compared to planets on wider orbits \citep{Dai2018}. These observations suggest USPs have experienced inclination excitation and orbital shrinkage, which may indicate the existence of additional, non-transiting companions. USPs in multi-planet systems often have formation scenarios invoking perturbers, so radial velocity (RV) or astrometric observations of such systems are important to place a constraint on additional, non-transiting planets. These well-characterized USP systems will be required to confirm and refine any USP formation scenarios.

In this paper, we investigate the M dwarf system, KOI-4777 ($V=16.4$, $J=13.2$), hosting a USP that was initially classified as a false positive by the Kepler DR25 automatic vetting. We validate the planetary nature of the Mars-sized ($R=0.51 \pm 0.03 R_\oplus$) transiting companion. We use publicly available observations along with precision near-infrared (NIR) RVs with the Habitable-zone Planet Finder Spectrograph \citep[HPF;][]{Mahadevan2012,Mahadevan2014} for statistical validation and to constrain the presence of non-transiting planets.

This paper is structured as follows. Section \ref{sec:obs} presents the observations used in this paper and Section \ref{sec:fpp} describes the false positive analysis of KOI-4777.01 using the \texttt{VESPA} statistical validation tool \citep{Morton2012,Morton2015}. Section \ref{sec:stellarpar} describes the method for spectroscopic characterization and our best estimates of the stellar parameters. In Section \ref{sec:tranfit}, we explain the analysis of the photometric and RV data while Section \ref{sec:disc} provides further discussion of the bulk properties of the KOI-4777 system and the feasibility for future study through additional high-precision RV observations. Finally, we conclude the paper in Section \ref{sec:summary} with a summary of our key results.

\section{Observations}\label{sec:obs}
\subsection{Photometry}
Kepler observed KOI-4777 (KIC 6592335, Gaia EDR3 2102436351673656704) for the entirety of the original mission in long-cadence mode with data from 2009 May 13 through 2013 May 11. It is not included in the final catalog (DR25) released by the Kepler team, which uses a fully automated vetting pipeline \citep{Coughlin2016,Mullally2016,Twicken2016} to catalog genuine transit events, or Kepler objects of interest (KOIs), in a uniform manner to maximize the reliability of the final catalog. The default search in the Kepler vetting pipeline did not consider transits with periods \(<0.5\) days \citep{Coughlin2019} and KOI-4777.01 was reported to have a period of 0.824 days, twice the true orbital period, and subsequently classified as a false positive. Manual vetting by members of the Kepler False Positive Working Group (FPWG) determined this system was a small planetary candidate with a period of 0.412 days. KOI-4777 is correctly identified in the supplemental Kepler DR25 candidate list as KOI-4777.01 \citep{Thompson2018}. 

\begin{figure*}[!ht]
\epsscale{1.15}
\plotone{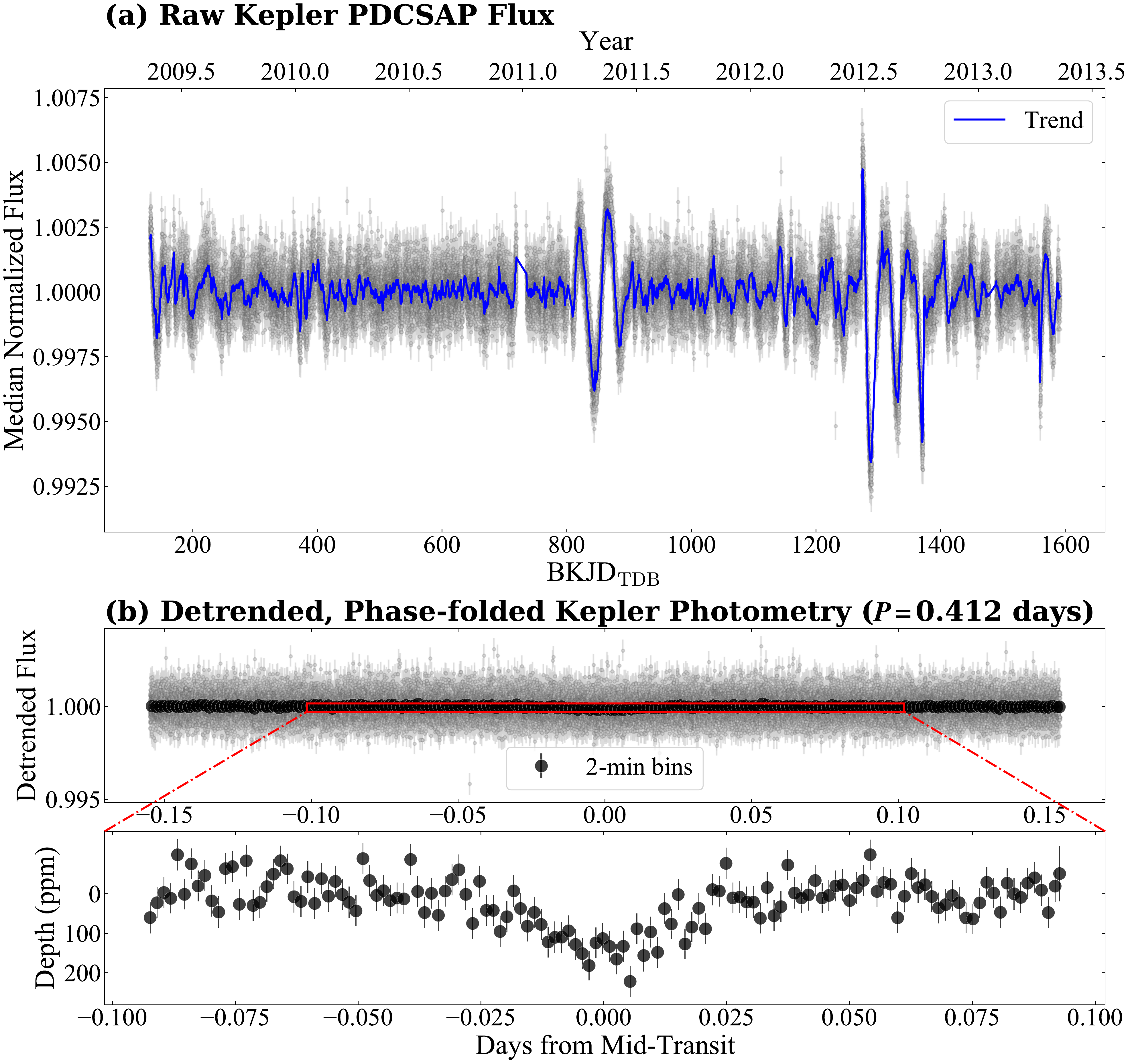}
\caption{Kepler Photometry of KOI-4777. \textbf{(a)} displays the raw PDCSAP photometry from all 17 quarters of the Kepler mission. We have excluded observations with non-zero data quality flags. \textbf{(b)} shows the photometry after detrending with a Gaussian process and phasing to the observed ephemeris from the supplemental Kepler DR 25 catalog. In the bottom panel, we bin the phase-folded data into 2-min bins to show the shape of the transit.}
\label{fig:4777phot}
\end{figure*}

For our subsequent analysis, we used the entire pre-search data-conditioned \citep[PDCSAP;][]{Stumpe2012,Smith2012} light curves available at the Mikulski Archive for Space Telescopes (MAST). We use the PDCSAP light curves from all 17 quarters of Kepler and exclude observations with non-zero data quality flags. These flags indicate poor-quality data due to conditions such as spacecraft events or cosmic ray hits and are described in the Kepler Archive Manual \citep[see Table 2-3 in][]{Thompson2016}. We do not perform additional processing or apply outlier rejection beyond the data quality flags. The raw photometry and the candidate signal are shown in Figure \ref{fig:4777phot}.

The true period was independently identified by \cite{Caceres2019} in the application of the autoregressive planet search procedure \citep[ARPS;][]{Caceres2019a} to 156,717 Kepler PDCSAP light curves. The ARPS analysis has four stages: (i) fitting and removing non-stationary stellar variations with low-dimensional autoregressive integrated moving average (ARIMA) models, (ii) calculating a periodogram using the transit comb filter (TCF) for the residuals, (iii) classifying light curve and TCF features with a random forest classifier trained on confirmed Kepler planet candidates, and (iv) vetting to remove false alarms and false positives. The fit removed nearly all of the variability seen in Figure \hyperref[fig:4777phot]{1(a)}.

\begin{figure*}[!ht]
\epsscale{1.15}
\plotone{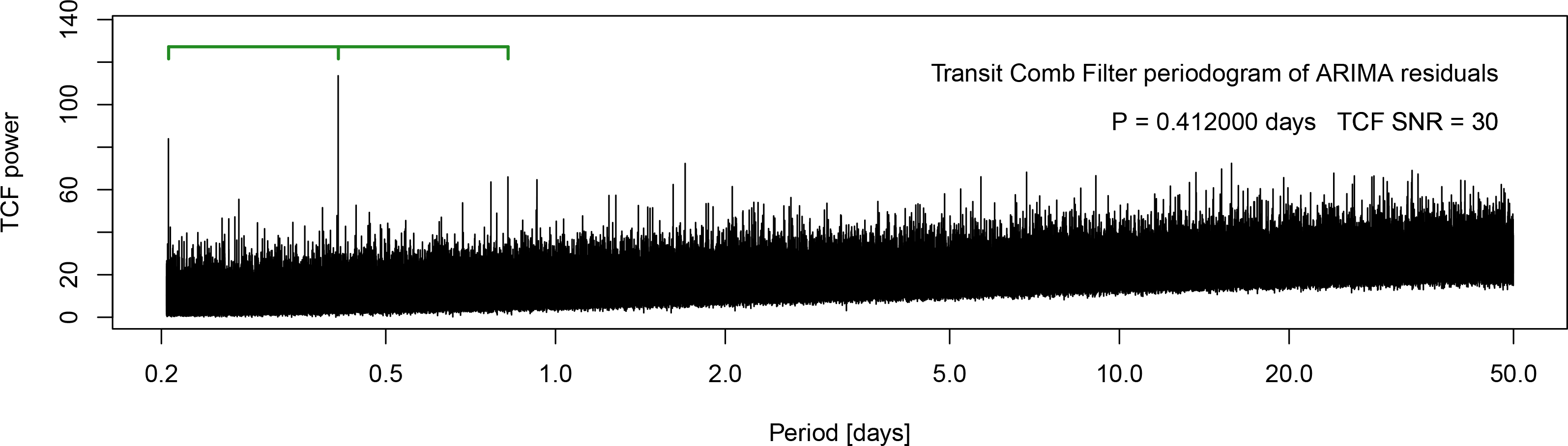}
\caption{Detection of KOI-4777.01 with ARPS. The periodogram is displayed from 0.2 to 50 days of KOI-4777.01 (KIC 6592335) obtained with the transit comb filter method in the ARPS analysis of Kepler light curves \citep{Caceres2019a,Caceres2019}. The strongest peak is at $P=0.412$ days and its harmonics are marked in green.}
\label{fig:4777arps}
\end{figure*}

The TCF periodogram of the residuals from the ARPS analysis for KOI-4777 is shown in Figure \ref{fig:4777arps} and reveals a periodic signal at $P=0.412$ days with S/N$=$30; weaker harmonics at 0.206 and 0.824 days are marked for reference. The transit depth estimated from the TCF matched filter is \(\sim126\) ppm with respect to the time-averaged median flux in the full Kepler light curve with an approximate transit duration of 0.5 hours. The random forest classifier, using several dozen features from different stages of the ARPS analysis, gives a high probability of planetary transit origin with P$_{RF}=0.64$, considerably above the P$_{RF}=0.35$ threshold chosen by \cite{Caceres2019} to classify a transit as a planet candidate.

The detection of a 0.412-day transit-like periodicity in KOI-4777.01 is designated KACT 39 by \cite{Caceres2019} in their list of 97 Kepler ARPS candidate planets. KOI-4777.01 has the second-highest random forest probability out of the 29 USP candidates. About 20 other KACT candidates have TCF spectral peaks with SNR$\geq$30 with  periods ranging from 0.2 to several days. KOI-4777.01 was also one of four objects where the DR25 KOI period was a long-period alias of the ARPS derived period, as the range of periods examined by the Kepler team excluded the true period. 

\subsection{High-resolution Doppler Spectroscopy}
We obtained fifteen 945-second visits of KOI-4777 with HPF at a median signal-to-noise ratio (S/N) per 1D extracted pixel of 19 at $1000\unit{nm}$. The HPF exposure time calculator\footnote{\url{https://psuastro.github.io/HPF/Exposure-Times/\#estimating-hpf-exposure-times}} suggests a nominal observation of a star with the same $J$ magnitude as KOI-4777 would have a median S/N per pixel of 25. HPF is a high-resolution ($R\sim55,000$), NIR (\(8080-12780\)\ \AA) spectrograph located at the 10m Hobby-Eberly Telescope (HET) at McDonald Observatory in Texas \citep{Mahadevan2012,Mahadevan2014} that achieves a long-term temperature stability of $\sim$1$\unit{mK}$ \citep{stefansson2016}. Our observations span almost a year from 20 June 2019 through 31 July 2020 and were executed in a queue by the HET resident astronomers \citep{Shetrone2007}. We use the algorithms in the tool \texttt{HxRGproc} for bias noise removal, nonlinearity correction, cosmic-ray correction, and slope/flux and variance image calculation \citep{Ninan2018} of the raw HPF data. The one dimensional spectra are reduced using the procedures in \cite{Ninan2018}, \cite{Kaplan2019}, and \cite{Metcalf2019}. 

HPF has a NIR laser frequency comb (LFC) calibrator to provide a precise wavelength solution and track instrumental drifts \citep{Metcalf2019}. We do not use simultaneous LFC calibrations during the observations to minimize the risk of contaminating our faint target spectrum with scattered light from the LFC and instead extrapolate the wavelength solution from LFC frames taken as part of standard evening/morning calibrations and from LFC calibration frames that are taken periodically throughout the night. The extrapolation from LFC frames enables precise wavelength calibration on the order of $<30 \unit{cm/s}$ \citep{Stefansson2020}, a value much smaller than the RV uncertainty for a faint target like KOI-4777. 

The RVs are derived following the methodology described in \cite{Stefansson2020}. Briefly, we use a modified version of the \texttt{SpEctrum Radial Velocity AnaLyser} pipeline \citep[\texttt{SERVAL};][]{Zechmeister2018}, which employs the template-matching technique \citep[e.g.,][]{Anglada-Escude2012} to derive RVs. \texttt{SERVAL} creates a master template from the observations and determines the Doppler shift for each individual spectrum by minimizing the \(\chi^2\) statistic. It generates the master template using all observed spectra for KOI-4777 after excluding regions with significant telluric contamination that are flagged with a synthetic telluric-line mask generated from \texttt{telfit} \citep{Gullikson2014}, a Python wrapper to the Line-by-Line Radiative Transfer Model package \citep{clough2005}. \texttt{SERVAL} calculates the barycentric correction for each epoch using \texttt{barycorrpy} \citep{Kanodia2018}, which uses the algorithms from \cite{wright2014}. Table \ref{tab:koirvs} presents the derived RVs, the \(1\sigma\) uncertainties, and the S/N per pixel at 1000 nm for KOI-4777.

\begin{deluxetable}{lrcc}
\tablecaption{RVs of KOI-4777.\tablenotemark{a} \label{tab:koirvs}}
\tablehead{\colhead{$\unit{BJD_{TDB}}$}  &  \colhead{RV}                   & \colhead{$\sigma$}  & \colhead{S/N \tablenotemark{b}} \\
           \colhead{}                    &  \colhead{$(\unit{m/s})$} & \colhead{$(\unit{m/s})$}  & \colhead{$@1000 \unit{nm}$} }
\startdata
2458654.956552 &     9 &   66 & 27 \\
2458654.967588 &   213 &   92 & 27 \\
2458656.941803 &   -68 &  131 & 19 \\
2458656.953302 &  -120 &  103 & 19 \\
2459004.772791 &   -28 &  140 & 16 \\
2459004.783857 &   380 &  140 & 16 \\
2459005.760007 &   -23 &  230 & 10 \\
2459026.702878 &    21 &  152 & 18 \\
2459026.714081 &     8 &  107 & 18 \\
2459030.686671 &  -293 &  132 & 19 \\
2459030.698159 &    61 &  100 & 19 \\
2459045.652346 &    83 &  120 & 21 \\
2459045.663519 &    29 &   93 & 21 \\
2459061.845708 &   -10 &   85 & 23 \\
2459061.857467 &    29 &   95 & 23 \\
\enddata
\tablenotetext{a}{All exposure times are 945s.}
\tablenotetext{b}{Per pixel}
\end{deluxetable}

\subsection{Sky-Projected Companions}
\begin{figure*}[t]
\epsscale{1.15}
\plotone{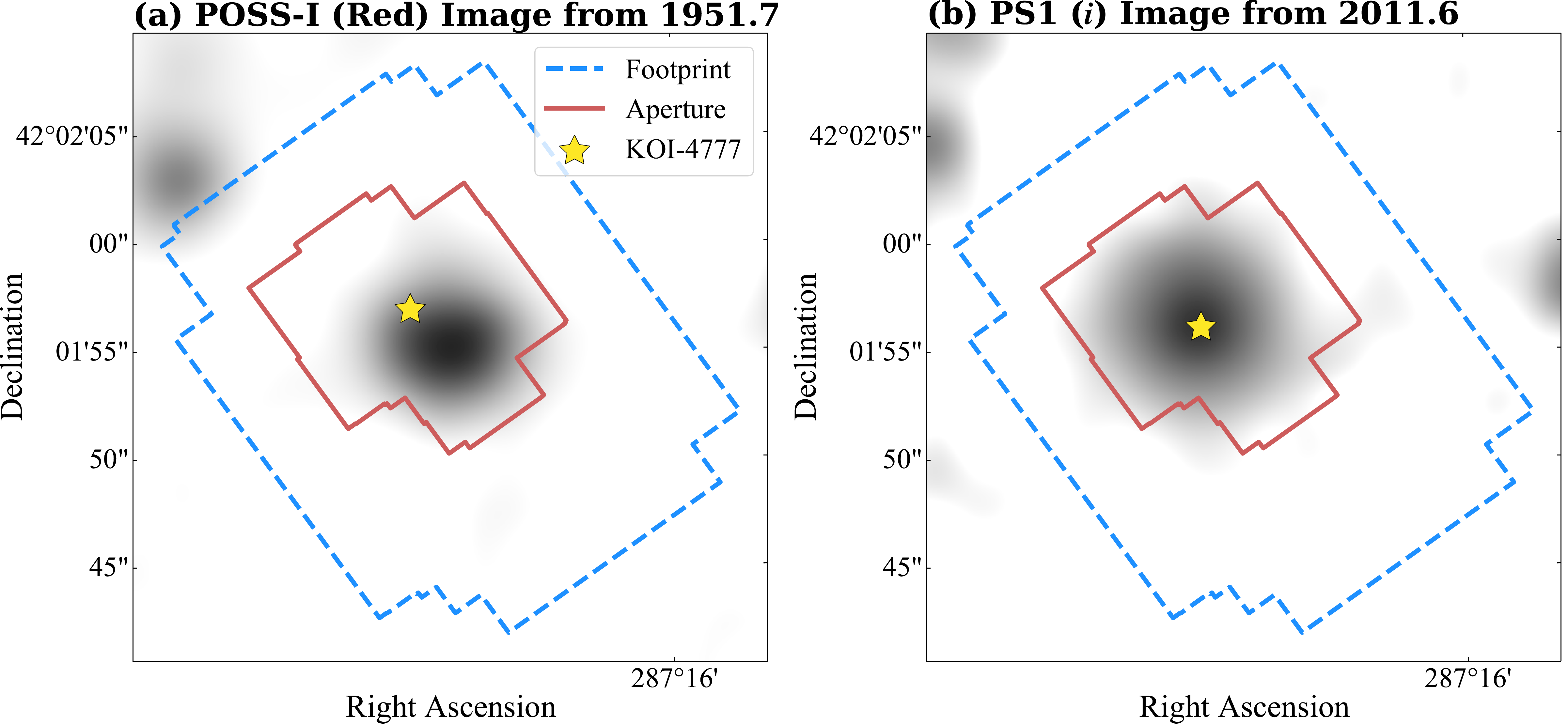}
\caption{Stellar neighborhood around KOI-4777. \textbf{(a)} presents the cumulative Kepler footprint on a POSS-I red image from 1951. The blue dashed polygon is the region of the sky observed at least once in all seventeen quarters of the Kepler data. The red solid polygon is the cumulative region of sky contained at least once within the aperture pixel mask. KOI-4777 is marked as a star. Gaia EDR3 detects no bright \(\Delta G<4\) companions within the entire Kepler footprint. \textbf{(b)} is similar to Panel A but overlaid on a PS1 \(i\) image taken in 2011 when Kepler was observing this star. KOI-4777 has moved slightly but no other bright stars have moved into the aperture or Kepler footprint. When KOI-4777 was observed by Kepler, there were no bright stars in the aperture that could cause significant dilution.}
\label{fig:4777pixel}
\end{figure*}

To investigate the presence of background companions at separations >4'' from KOI-4777, we searched the entire region around KOI-4777 observed by Kepler using archival photometry and Gaia EDR3 \citep{GaiaCollaboration2021}. We use the Kepler target pixel files \citep[TPFs;][]{Kinemuchi2012} available on MAST to determine the extent of the sky contained in the Kepler footprint and obtain the optimal aperture used to generate the PDCSAP light curve. Figure \ref{fig:4777pixel} shows the TPF region for KOI-4777 and the respective apertures: the region of sky observed at least once by Kepler is indicated the dashed polygon while the region included in the optimal aperture masks is marked as a solid polygon.

Gaia EDR3 indicates that only KOI-4777 is contained in the Kepler footprint and in the optimal aperture. The region of sky around the Kepler footprint was observed by the first Palomar Sky Survey \citep[POSS-I;][]{Minkowski1963} in 1951 and the Pan-STARRS1 Surveys \citep[PS1;][]{Chambers2016,Magnier2020} in 2011. Figure \hyperref[fig:4777pixel]{3(a)} presents the POSS-I red image overlaid with the Kepler footprint and aperture. KOI-4777 is off-center in the aperture and there are no additional bright background sources contained the aperture or footprint. Figure \hyperref[fig:4777pixel]{3(b)} displays the PS1 $i$ image of KOI-4777 with an identical overlay showing KOI-4777 is centered on the Kepler aperture and no other background stars have entered the Kepler footprint. KOI-4777 has no detectable bright on-sky companions in either Gaia EDR3 or archive imaging such that the Kepler light curve should not contain any significant dilution due to contaminating light from other stars.

\subsection{High-Contrast Imaging}
\begin{figure*}[!ht]
\epsscale{1.15}
\plotone{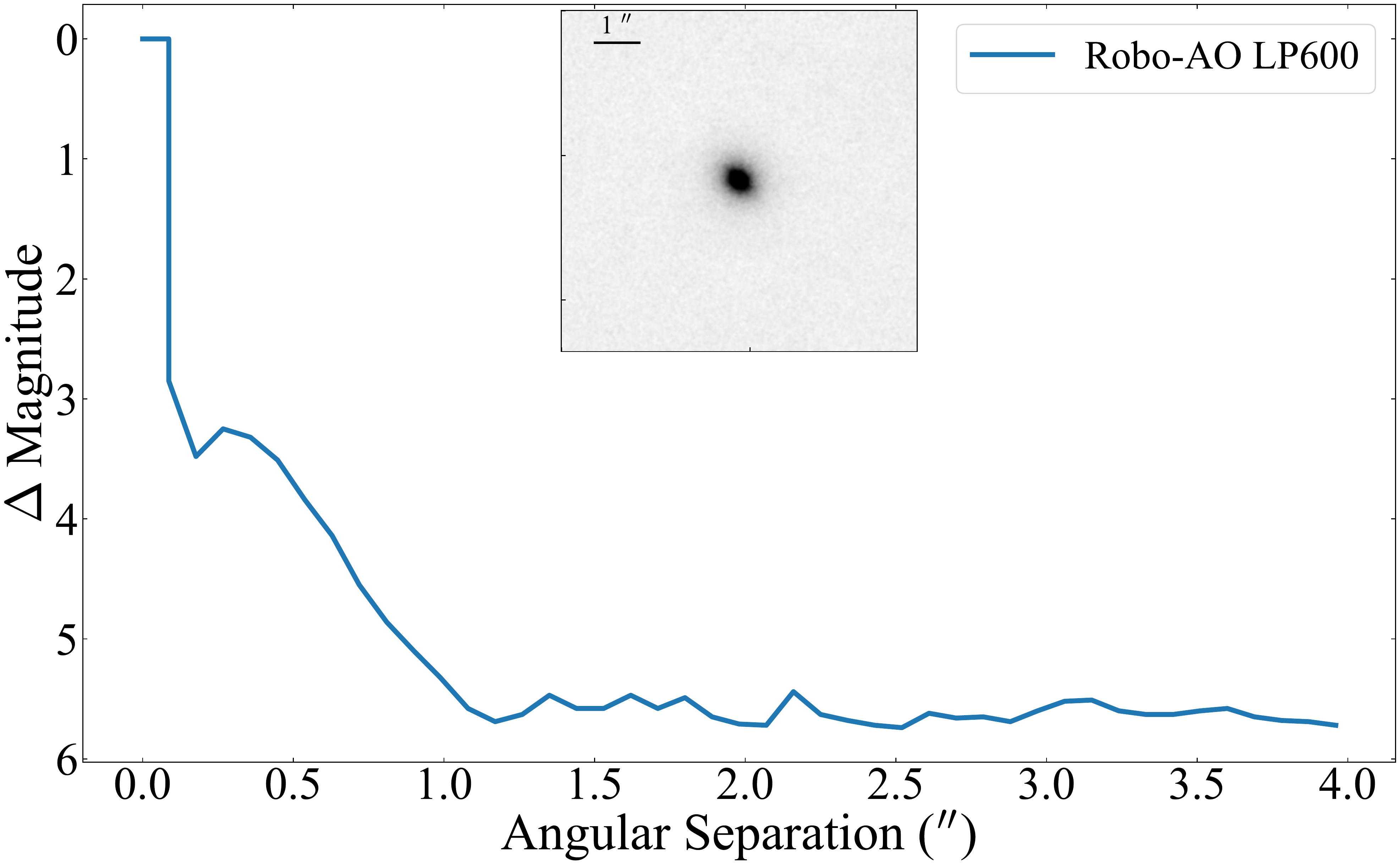}
\caption{Robo-AO Imaging of KOI-4777. This figure displays the \(5\sigma\) contrast curve observed by \cite{Ziegler2018a} using Robo-AO in a long-pass filter with a hard cut off at 600 nm (LP-600). The data show there are no bright companions within 4\arcsec{} of the host star. The inset image is an 8\arcsec cutout centered on KOI-4777 in this long-pass filter. The contrast curve and image were obtained from the Robo-AO KOI survey (\url{http://roboaokepler.org/koi_pages/KOI-4777.html}).}
\label{fig:4777ao}
\end{figure*}

KOI-4777 was observed as part of the Robo-AO Kepler planetary candidate survey \citep{Ziegler2018a} on 19 June 2016. The observations were performed using the Robo-AO laser adaptive optics system \citep{Baranec2013,Baranec2014} on the 2.1-m telescope at Kitt Peak National Observatory \citep{Jensen-Clem2018} using a 1.85-m circular aperture mask on the primary mirror. These observations were taken using a long-pass filter with a hard cut off at 600 nm that was designed by the Robo-AO team to approximate the Kepler bandpass at redder wavelengths and suppress blue wavelengths to minimize the impact on adaptive optics performance. The resulting $5\sigma$ contrast curve is shown in Figure \ref{fig:4777ao}. The Robo-AO observations reveal that are no bright ($\Delta \mathrm{mag}$ < 4) secondary companions within $4.0\arcsec$. 

\section{False Positive Probability Analysis}\label{sec:fpp}
The shallow depth reported by the Kepler DR25 supplemental catalog (121 ppm) prevents additional ground-based observations of a transit of KOI-4777.01. We instead employ the package Validation of Exoplanet Signals using a Probabilistic Algorithm \citep[\texttt{VESPA};][]{Morton2016} to conduct a false-positive analysis of KOI-4777.01. The algorithm implements the statistical techniques described in \cite{Morton2012} to validate a planet by simulating and determining the likelihood of a range of astrophysical false-positive scenarios, including background eclipsing binaries (BEBs), eclipsing binaries (EBs), and hierarchical eclipsing binaries (HEBs). \texttt{VESPA} generates a population for each scenario and calculate the respective likelihood. KOI-4777.01 was previously analyzed by \texttt{VESPA} in \cite{Morton2016}, albeit at twice the orbital period, where it was determined to have an FPP of 0.96 and most likely a background eclipsing binary. 

We update the \texttt{VESPA} analysis with the period of \(P=0.412\) days \citep{Coughlin2019,Caceres2019a}, and include the contrast curve from Robo-AO as an additional constraint to limit the brightness of any background companions within 4\arcsec of KOI-4777. As inputs to \texttt{VESPA}, we use the i) segment of the phase-folded Kepler transit centered on the transit and buffered by a baseline three times the transit duration, ii) 2MASS $J$, $H$, $K$ and Kepler magnitudes, iii) Gaia EDR3 parallax, iv) host star stellar effective temperature, surface gravity, and metallicity. We set a uniform prior on the visual extinction where the upper limit is determined using estimates of Galactic dust extinction by \cite{Green2019}. The maximum radius permissible for a BEB is the upper \(3\sigma\) centroid offset determined by the Kepler DR 25 pipeline (3.3169\arcsec) and the maximum depth of the secondary transit is the rms of the light curve after excising the transits (\(<190\) ppm). We include the Robo-AO contrast curves shown in Figure \ref{fig:4777ao} as a constraint applied to the BEB population during the \texttt{VESPA} analysis.

The shorter period used in this analysis meant that all EB systems at a period of 0.412 days would exceed the Roche limit of the host star \citep[see Equation 2 in][]{Eggleton1983}, such that the resulting system would be a contact or semi-detached binary and show a different morphology in the light curve than what is observed by Kepler. For our analysis, \texttt{VESPA} only considered the FPP contribution from double-period EB and HEB systems. 

We obtain an FPP of \(0.008\pm0.001\) for KOI-4777.01 from 100 bootstrap recalculations of the initially simulated populations for KOI-4777. Similar to the analysis by \cite{Morton2016}, BEBs are the dominant FPP scenario and, adopting the threshold of \(0.01\) from \cite{Morton2016}, KOI-4777 may be considered a validated planet. 

\section{Stellar Parameters}\label{sec:stellarpar}
\subsection{Spectroscopic Parameters}
To derive spectroscopic stellar parameters of KOI-4777, we use \texttt{HPF-SpecMatch} \cite[][Jones et. al. 2021 in prep]{Stefansson2020} which is based on the methodology discussed in \cite{Yee2017}. \texttt{HPF-SpecMatch} derives the stellar properties of KOI-4777 by comparing the highest S/N HPF spectra of KOI-4777 to a library of 86 high quality (\(\unit{S/N}>100\)) HPF stellar spectra with well-determined properties \citep[see][]{Yee2017} spanning: $3000 \unit{K} < T_{e} < 5500 \unit{K}$, $4.4<\log g < 5.2$, and $-0.5 < \mathrm{[Fe/H]} < 0.5$.

For this analysis, \texttt{HPF-SpecMatch} compares the spectral order containing $8670-8750$ \AA\ to the HPF spectral library because there is minimal telluric contamination in the $z$-band. The algorithm identifies the best-matching library spectrum using \(\chi^{2}\) minimization,  creates a composite spectrum from a weighted linear combination of the five best-matching library spectra, and derives the stellar properties using the calculated weights. The uncertainty for each stellar parameter ($T_e$, $\log g$, and [Fe/H]) is the standard deviation of the residuals from a leave-one-out cross-validation procedure applied to the entire stellar library in this wavelength region. The derived parameters for KOI-4777 are $T_{\mathrm{eff}}=3515 \pm 69 \unit{K}$, $\mathrm{[Fe/H]} = 0.1 \pm 0.1$ and $\log (g) = 4.77 \pm 0.04$ and are listed in Table \ref{tab:stellarparam}. KOI-4777 has an approximate M1.5 spectral type when using the $T_{e}$ classification from Table 2 of \cite{Worthey1994}.

\startlongtable
\begin{deluxetable*}{lccc}
\tabletypesize{\scriptsize}
\tablecaption{Summary of Stellar Parameters. \label{tab:stellarparam}}
\tablehead{\colhead{~~~Parameter}&  \colhead{Description}&
\colhead{Value}&
\colhead{Reference}}
\startdata
\multicolumn{4}{l}{\hspace{-0.2cm} Main identifiers:}  \\
~~~KIC &  \(\cdots\)  & 6592335 & KIC \\
~~~KACT &  \(\cdots\)  & 39 & ARPS \\
~~~Gaia EDR3 & \(\cdots\) & 2102436351673656704 & Gaia EDR3 \\
\multicolumn{4}{l}{\hspace{-0.2cm} Equatorial Coordinates, Proper Motion, Distance, and Extinction:} \\
~~~$\alpha_{\mathrm{J2016}}$ &  Right Ascension (RA) & 19:09:02.92 & Gaia EDR3 \\
~~~$\delta_{\mathrm{J2016}}$ &  Declination (Dec) & 42:01:56.01 & Gaia EDR3 \\
~~~$l$ &  Galactic Longitude & 72.95336 & Gaia EDR3 \\
~~~$b$ &  Galactic Latitude & 14.74734 & Gaia EDR3 \\
~~~$\mu_{\alpha}$ &  Proper motion (RA, \unit{mas/yr}) & $-2.149\pm0.029$ & Gaia EDR3 \\
~~~$\mu_{\delta}$ &  Proper motion (Dec, \unit{mas/yr}) & $-13.853\pm0.032$ & Gaia EDR3  \\
~~~$d$ &  Distance in pc$^a$  & $170.66_{-0.74}^{+0.70}$ & Bailer-Jones\\
~~~\(A_{V,max}\) & Maximum visual extinction & $0.03$ & Green\\
\multicolumn{4}{l}{\hspace{-0.2cm} Optical and near-infrared magnitudes:}  \\
~~~$B$ & Johnson $B$ mag & $18.305\pm0.034$ & EHK\\
~~~$V$ & Johnson $V$ mag & $16.921\pm0.022$ & EHK\\
~~~$J$ & 2MASS $J$ mag & $13.221\pm0.021$ & 2MASS\\
~~~$H$ & 2MASS $H$ mag & $12.671\pm0.021$ & 2MASS\\
~~~$K_s$ & 2MASS $K_s$ mag & $12.449\pm0.018$ & 2MASS\\
~~~$W1$ &  WISE1 mag & $12.331\pm0.023$ & WISE\\
~~~$W2$ &  WISE2 mag & $12.210\pm0.022$ & WISE\\
~~~$W3$ &  WISE3 mag & $12.288\pm0.274$ & WISE\\
\multicolumn{4}{l}{\hspace{-0.2cm} Spectroscopic Parameters$^b$:}\\
~~~$T_{e}$ &  Effective temperature in \unit{K} & $3515 \pm 69$& This work\\
~~~$\mathrm{[Fe/H]}$ &  Metallicity in dex & $0.1\pm0.1$ & This work\\
~~~$\log(g)$ & Surface gravity in cgs units & $4.77 \pm 0.04$ & This work\\
\multicolumn{4}{l}{\hspace{-0.2cm} Model-Dependent Stellar SED and Isochrone fit Parameters$^c$:}\\
~~~$M_*$ &  Mass in $M_{\odot}$ & $0.41\pm0.02$ & This work\\
~~~$R_*$ &  Radius in $R_{\odot}$ & $0.40\pm0.01$ & This work\\
~~~$\rho_*$ &  Density in $\unit{g/cm^{3}}$ & $8.9\pm0.6$ & This work\\
~~~$A_v$ & Visual extinction in mag & $0.013\pm0.009$ & This work\\
\multicolumn{4}{l}{\hspace{-0.2cm} Other Stellar Parameters:}           \\
~~~$v \sin i_*$ &  Rotational velocity in \unit{km/s}  & $<2$ & This work\\
~~~$P_{rot}$ &  Rotational period in days  & $44\pm1$ & This work\\
~~~Age & Age in Gyrs & $2-5$ & This work\\
~~~$RV$ &  Radial velocity in \unit{km/s} & $28.11\pm0.05$ & This work\\
~~~$U, V, W$ &  Barycentric Galactic velocities in \unit{km/s} &  $18.72\pm0.06, 23.46\pm0.05, 4.38\pm0.03$ & This work\\
~~~$U_{\mathrm{LSR}}, V_{\mathrm{LSR}}, W_{\mathrm{LSR}}$ &  Galactic velocities w.r.t. LSR$^d$ in \unit{km/s} &  $29.8\pm0.8, 35.7\pm0.5, 11.6\pm0.4$ & This work\\
\enddata
\tablenotetext{}{References are: KIC \citep{Brown2011}, ARPS \citep{Caceres2019a}, Gaia EDR3 \citep{GaiaCollaboration2021}, Bailer-Jones \citep{Bailer-Jones2021}, Green \citep{Green2019}, EHK \citep{Everett2012}, 2MASS \citep{Cutri2003}, and WISE \citep{Wright2010}}
\tablenotetext{a}{Geometric distance from \cite{Bailer-Jones2021}.}
\tablenotetext{b}{Derived using our modified \texttt{HPF-SpecMatch} algorithm.}
\tablenotetext{c}{\texttt{EXOFASTv2} derived values using MIST isochrones.}
\tablenotetext{d}{Calculated using the solar velocities from \cite{Schoenrich2010}.}
\end{deluxetable*}

We use \texttt{galpy} \citep{Bovy2015} to calculate the \textit{UVW} velocities in the barycentric frame using the Gaia EDR3 proper motions and the systemic velocity derived from HPF. The reported values are in a right-handed coordinate system \citep{Johnson1987} such that \textit{UVW} are positive in the directions of the Galactic center, Galactic rotation, and the north Galactic pole, respectively. The \(UVW\) velocities are calculated with respect to the local standard of rest using the solar velocities and uncertainties from \cite{Schoenrich2010}. KOI-4777 is classified as a field star in the thin disk after applying the kinematic selection criteria from \cite{Bensby2014} to the derived \(UVW\) velocities. The BANYAN \(\Sigma\) algorithm \citep{Gagne2018}, which derives cluster membership probabilities using sky positions, proper motions, parallax, radial velocities, and spectrophotometric distance constraints, further classifies KOI-4777 as a field star that is not associated with any young clusters. 

\subsection{Spectral Energy Distribution Fitting}
We use the {\tt EXOFASTv2} analysis package \citep{Eastman2019} to model the spectral energy distribution (SED) and derive model-dependent stellar parameters using the MIST stellar models \citep{Dotter2016,Choi2016}. {\tt EXOFASTv2} calculates the bolometric corrections for the SED fit by linearly interpolating the precomputed bolometric corrections supplied by the MIST team in a grid of \(\log g\), \(\mathrm{T_{eff}}\), [Fe/H], and \(A_V\)\footnote{\url{http://waps.cfa.harvard.edu/MIST/model_grids.html\#bolometric}}. The MIST grid is based on the ATLAS12/SYNTHE stellar atmospheres \citep{Kurucz1970,Kurucz1993}. The fit uses Gaussian priors on the (i) 2MASS \(JHK\) magnitudes, Johnson \(BV\) magnitudes from \cite{Everett2012}, and Wide-field Infrared Survey Explorer magnitudes \citep{Wright2010}; (ii) host star surface gravity, temperature, and metallicity derived with \texttt{HPF-SpecMatch}; and (iii) the geometric distance estimate from \cite{Bailer-Jones2021} and a uniform prior for the visual extinction in which the upper limit is determined from estimates of Galactic dust by \cite{Green2019} calculated at the distance determined by \cite{Bailer-Jones2021}. The \(R_{v}=3.1\) reddening law from \cite{Fitzpatrick1999} is used by \texttt{EXOFASTv2} to convert the extinction determined by \cite{Green2019} to a visual magnitude extinction. The derived stellar parameters with their uncertainties are listed in Table \ref{tab:stellarparam}. We derive a mass and radius of \(0.41\pm0.02~\mathrm{M_{\odot}}\) and \(0.40\pm0.01~\mathrm{R_{\odot}}\) for KOI-4777. 

\subsection{Rotation period \& constraints on system age} \label{sec:rot}
The rotation period and kinematics for M dwarfs can provide an estimate of the age for the star because rapidly-rotating M dwarfs are typically younger than slowly-rotating counterparts \citep[e.g.,][]{Irwin2011,Newton2016}. Our analysis with \texttt{HPF-SpecMatch} also broadens the spectra using a linear limb darkening law \citep{Yee2017} when generating the best-fitting template. Using our HPF spectra, we can only place a constraint of $v \sin i < 2 \unit{km/s}$ due to the resolution of $\sim 55,000$. The activity indicators from HPF spectra \citep[see][]{Zechmeister2018,Stefansson2020b}, including the differential line width, the chromatic RV index, and activity indices of the three lines in the Calcium II infrared triplet, show no variability and cannot be used to gauge the rotation period of the star. This suggests that KOI-4777 is probably not young and probably has a long rotation-period. 

We use the available Kepler photometry to confirm the existence of a long rotation period. We do not search for a rotation period in the PDCSAP data (shown in \ref{fig:4777phot}) because the Kepler team warns that long period signals are attenuated in the PDCSAP flux and, in the latest iteration of the pipeline, the algorithm assumes all long period signals are systematics \citep[see Section 5.15 in][]{VanCleve2016}. \cite{Gilliland2015} showed that signals \(>20\) days are severely damped in the PDCSAP flux, although larger signals displayed comparatively better preservation at long periods. The Kepler team instead suggest searching for long-period signals by using the SAP flux and the available co-trending basis vectors \citep[e.g.,][]{Aigrain2017,Cui2019}. 

For this work, we use the \texttt{ARC2}\footnote{\url{https://github.com/OxES/OxKeplerSC}} pipeline developed by \cite{Aigrain2017} to correct for systematics in the Kepler SAP light curve. The \texttt{ARC2} pipeline performs the correction in two steps where it: (i) detects and removes isolated discontinuities from Kepler light curves (``jumps'' in the data) and (ii) removes the instrument systematic trends from the photometry by using the publicly available co-trending basis vectors \citep[CBVs; see][]{Kinemuchi2012}. \cite{Aigrain2017} note that the exact number of CBVs to use will vary, but the final correction is generally insensitive to the number of CBVs if more than 2-3 CBVs are used. To determine the number of CBVs to use for KOI-4777, we ran the \texttt{ARC2} pipeline with anywhere from 2-8 CBVs and compared the generalized Lomb-Scargle (GLS) periodogram \citep{Zechmeister2009} between 1-100 days for each detrended light curve. Using seven CBVs provided the highest power in the periodogram, and we adopt this as our detrended SAP photometry. All light curves derived from \texttt{ARC2} had a significant peak at \(\sim 43\) days regardless of the number of CBVs used. 

\begin{figure*}[!ht]
\epsscale{1.15}
\plotone{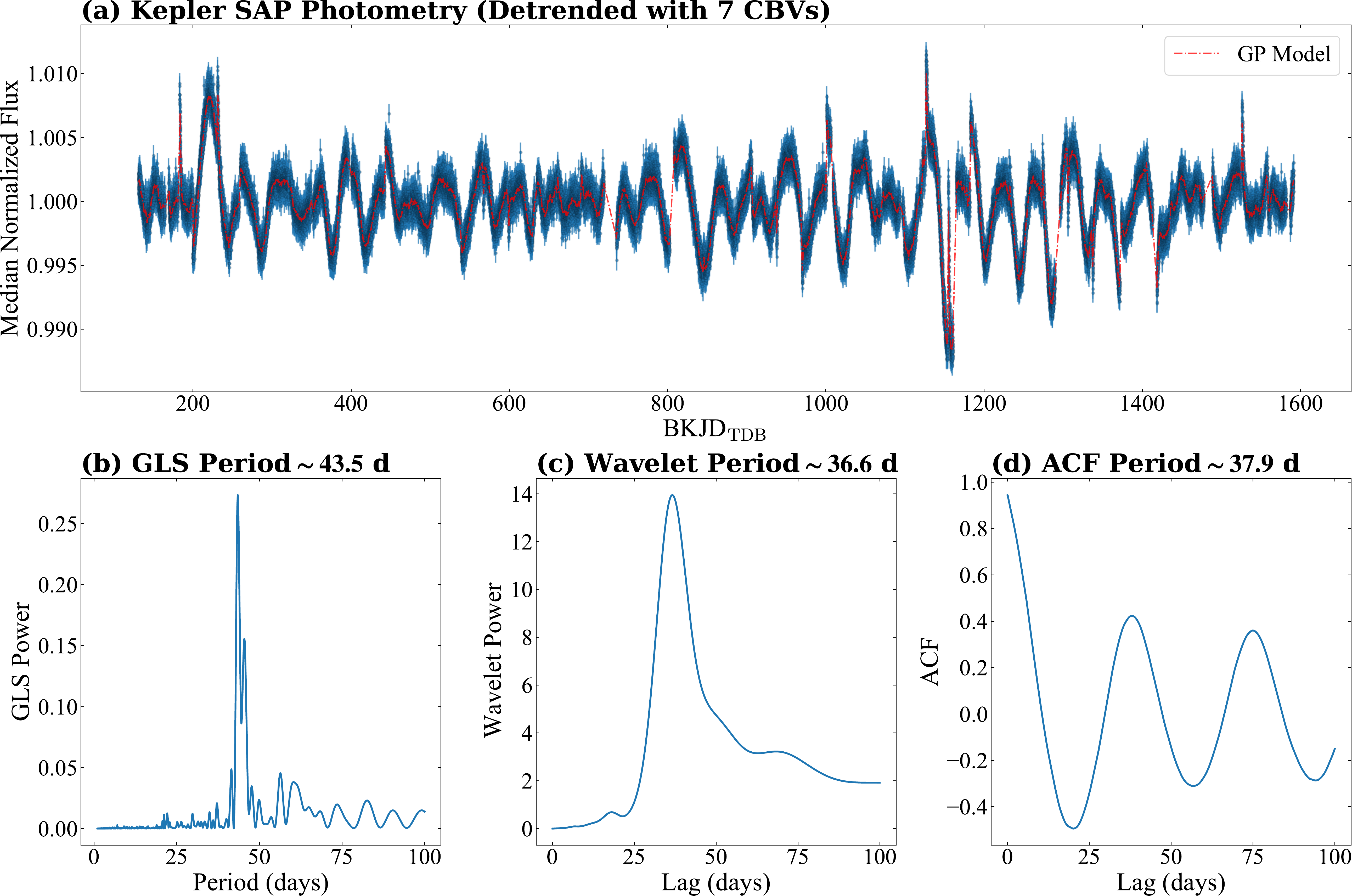}
\caption{Rotation period of KOI-4777. (a) displays the median-normalized Kepler SAP photometry, after detrending with seven CBVs, along with our Gaussian process (GP) model as the dash-dotted line. (b) shows the generalized Lomb-Scargle (GLS) periodogram, (c) the time lags derived via wavelet decomposition, and (d) shows the autocorrelation function of the photometry in panel (a). Each method in panels (b) - (d) shows a significant peak between 25 - 50 days.}
\label{fig:rotation}
\end{figure*}

We apply three common methods of time series analysis \citep[see][]{CantoMartins2020,Reinhold2020} to search for the rotation period in the detrended SAP flux: the GLS periodogram, the wavelet power spectrum \citep[e.g.,][]{Bravo2014}, and the auto-correlation function (ACF) \citep[e.g.,][]{McQuillan2013,McQuillan2013a}. All three methods indicate a rotation period in the range of 25-50 days, but the periods differ from each other due to the shape of the light curve and the non-uniformity of the sampling. To further constrain the rotation period, we employ the \texttt{juliet} analysis package \citep{Espinoza2019} to model the SAP photometry using the \texttt{celerite} package and the approximate quasi-periodic covariance function from \cite[][]{Foreman-Mackey2017}. This covariance function \citep[Equation 56 in][]{Foreman-Mackey2017} takes the form of 
\begin{equation}
k(\tau) = \frac{B}{2 + C} e^{-\tau/L} \left[ \cos \left( \frac{2 \pi \tau}{P_\mathrm{GP}} \right) + (1 + C) \right],
\label{eq:kernelperiodic}
\end{equation}
where $\tau$ is the scalar of choice (for KOI-4777, we use time), and $B$, $C$, $L$, and $P_{\mathrm{GP}}$ are the hyperparameters of the covariance function. $B$ and $C$ determine the weight of the exponential term with a decay constant of $L$ (in days). $P_{\mathrm{GP}}$ determines the periodicity of the quasi-periodic oscillations and is taken as an estimator of the stellar rotation period. Equation \ref{eq:kernelperiodic} has been shown to reproduce the behavior of a more traditional quasi-periodic covariance function and has allowed for computationally efficient inference of stellar rotation periods even for large datasets that are not uniformly sampled \citep[e.g.,][]{Angus2018}. We place a broad uniform prior on the rotation period of $1-100$ days. \texttt{juliet} performs the parameter estimation using \texttt{dynesty} \citep{Speagle2020}, a dynamic nested-sampling algorithm. Figure \ref{fig:rotation} displays the SAP photometry, the best fitting Gaussian process model, and the results from the GLS, wavelet transform, and the ACF. 

The Gaussian process modeling indicates a rotation period of $P_{\mathrm{rot}} = 44\pm1$ days. Our period estimate is consistent with the period of \(44.61\pm20.51\) days determined by \cite{McQuillan2013} using the first 4 quarters from Kepler. KOI-4777 has an intermediate rotation period using the classification criteria from \cite{Newton2016}, who were able to show that M dwarfs with $P_{\mathrm{rot}}<10$ days have a mean age of 2 Gyr while those with $P_{\mathrm{rot}}>70$ days have mean ages of $\sim 5$ Gyr. There is no age range for M dwarfs with rotation periods spanning 10–70 days in \cite{Newton2016} because this regime was sparsely populated. With this measured rotation period, KOI-4777 may have an age between $2-5$ Gyr. This range is compatible with our model-dependent SED fit and still allows for the existence of KOI-4777.01, as USPs have been shown to be stable against tidal inspiral during the lifetime of main sequence stars \citep{Hamer2020}.

\section{Transit and radial velocity modeling} \label{sec:tranfit}
We use \texttt{juliet} to jointly model the Kepler PDCSAP photometry and HPF RVs. \texttt{juliet} calculates the transit model with the \texttt{batman} package \citep{Kreidberg2015} and uniformly samples the limb-darkening parameters using the parameterization from \cite{Kipping2013b}. The transit model utilizes the supersampling option in \texttt{batman} with exposure times of 30 minutes and a supersampling factor of 30 due to the long-cadence of the Kepler data. To account for correlated noise in the Kepler PDCSAP photometry, the fit includes a Gaussian process noise model identical to the one discussed in Section \ref{sec:rot}. \texttt{juliet} models the RVs with a standard Keplerian RV curve generated from the \texttt{radvel} \citep{Fulton2018} package. Both the transit and RV models include a simple white-noise model in the form of a jitter term that is added in quadrature to the uncertainties. We adopt a circular orbit and force an eccentricity of \(e=0\) because the circularization timescale \citep[see][]{Goldreich1966,Jackson2008} for KOI-4777.01 is <100,000 years even if we conservatively adopt a tidal quality factor $Q_p=500$ for a terrestrial planet \citep[e.g.,][]{Jackson2008a} and a planetary mass of $1M_\oplus$. We also set a prior on the stellar density using the value determined from our \texttt{EXOFASTv2} SED fit. 

\begin{figure*}[!ht]
\epsscale{1.15}
\plotone{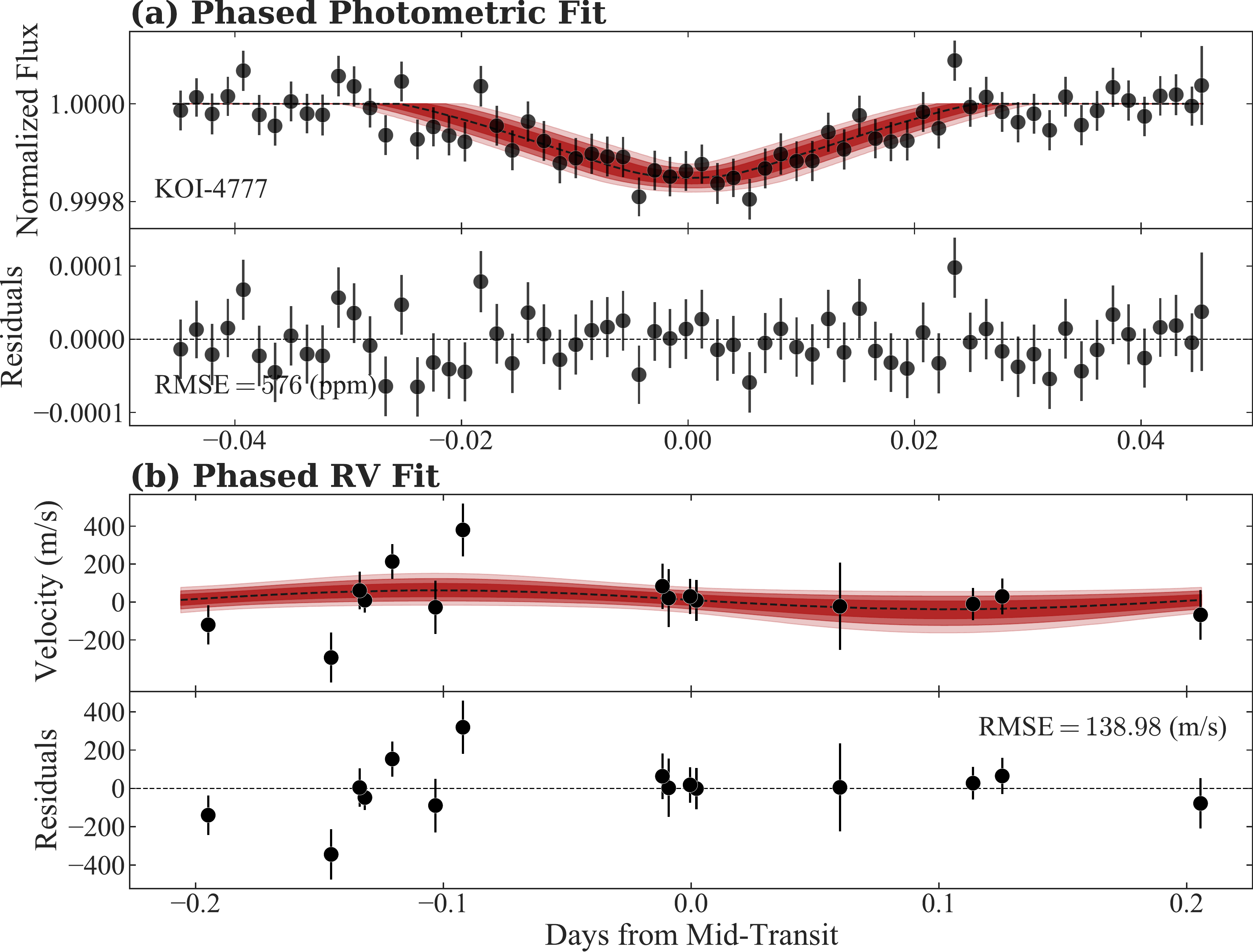}
\caption{\textbf{(a)} displays the photometric model and the phased Kepler light curve. For clarity, we only show the phase-folded data after rebinning to 2-min bins (similar to Figure \hyperref[fig:4777phot]{1(b)}). \textbf{(b)} presents the RVs after phasing the data to the ephemeris derived from the joint fit. For panels (a) and (b), the best-fitting model is plotted as a dashed line while the shaded regions denote the \(1\sigma\) (darkest), \(2\sigma\), and \(3\sigma\) range of the derived posterior solution.}
\label{fig:joint}
\end{figure*}

Figure \ref{fig:joint} presents the results of the fit and Table \ref{tab:derivedpar} provides a summary of the derived system parameters and respective confidence intervals. The joint fit to the photometric and spectroscopic data indicate KOI-4777.01 is a Mars-sized transiting companion ($0.51 \pm 0.03~ \mathrm{R}_{\oplus}$) on a \(0.412000 \pm 0.000001\) day orbit. These results are consistent to within their \(1\sigma\) uncertainties to values obtained from \texttt{EXOFASTv2} for a joint fit of the SED, HPF RVs, and the detrended photometry from the \texttt{juliet} fit. The HPF RVs place an upper mass constraint of $99.2 M_\oplus$ at the 99th quantile of the posterior distribution, which further strengthens the planetary validation because the observed transit cannot be due to an eclipsing stellar companion or a massive, grazing sub-stellar companion.

\startlongtable
\begin{deluxetable*}{llcc}
\tabletypesize{\scriptsize }
\tablecaption{Derived Parameters for KOI-4777 \label{tab:derivedpar}}
\tablehead{\colhead{~~~Parameter} &
\colhead{Units} &
\colhead{Prior} &
\colhead{Value}
}
\startdata
\sidehead{Transit Parameters:}
~~~Linear Limb-darkening Coefficient\dotfill & $q_1$\dotfill & $\mathcal{U}(0,1)^a$ & $0.3_{-0.2}^{+0.4}$ \\
~~~Quadratic Limb-darkening Coefficient\dotfill & $q_2$\dotfill & $\mathcal{U}(0,1)$ & $0.4 \pm 0.3$ \\
~~~Scaled Radius\dotfill & $R_{p}/R_{*}$ \dotfill & $\mathcal{U}(0,1)$ & $0.0116 \pm 0.0006$\\
~~~Impact Parameter\dotfill & $b$\dotfill & $\mathcal{U}(0,1)$ & $0.2_{-0.1}^{+0.2}$\\
~~~Photometric Jitter\dotfill & $\sigma_{\mathrm{Kepler}}$ (ppm)\dotfill & $\mathcal{J}(10^{-6},5000)^b$ & $165 \pm 6$\\
\sidehead{RV Parameters:}
~~~Orbital Period\dotfill & $P$ (days) \dotfill & $\mathcal{N}(0.412,0.01)^c$ & $0.412000 \pm 0.000001$\\
~~~Time of Conjunction\dotfill & $T_C$ (BJD\textsubscript{TDB})\dotfill & $\mathcal{N}(2454964.80421,0.01)$ & $2454965.219 \pm 0.002$\\
~~~Eccentricity\dotfill & $e$ \dotfill & Fixed & $0$\\
~~~Argument of Periastron\dotfill & $\omega$ (degrees) \dotfill & Fixed & $90$\\
~~~Semi-amplitude Velocity\dotfill & $K$ (m/s) \dotfill & $\mathcal{U}(0,10^3)$ & $3\sigma < 83$\\
~~~RV zeropoint \dotfill & $\gamma_{\mathrm{HPF}}$ (m/s) \dotfill & $\mathcal{U}(-10^{-3},10^3)$ & $10 \pm 30$\\
~~~RV jitter \dotfill & $\sigma_{\mathrm{HPF}}$ (m/s) \dotfill & $\mathcal{J}(10^{-3},10^3)$ & $0.6_{-0.6}^{+24.8}$\\
\sidehead{Gaussian Process Hyperparameters:}
~~~$B$\dotfill & Amplitude ($10^{-6} \unit{ppm}$)\dotfill& $\mathcal{J}(10^{-6},1)$ & $1.2_{-0.1}^{+0.2}$\\
~~~$C$\dotfill & Additive Factor (\(10^{-4}\)) \dotfill & $\mathcal{J}(10^{-6},10^6)$ & $0.000 \pm 0.007$\\
~~~$L$\dotfill & Length scale (days) \dotfill & $\mathcal{J}(10^{-6},10^6)$ & $27_{-3}^{+4}$\\
~~~$P_{GP}$\dotfill & Period (days) \dotfill & $\mathcal{J}(1,2000)$ & $37\pm2$\\
\sidehead{Derived Planetary Parameters:}
~~~Scaled Semi-major Axis\dotfill & $a/R_{*}$ \dotfill & $\cdots$ & $4.28 \pm 0.09$\\
~~~Orbital Inclination\dotfill & $i$ (degrees)\dotfill & $\cdots$ & $87 \pm 2$\\
~~~Transit Duration\dotfill & $T_{14}$ (hours)\dotfill & $\cdots$ & $0.73_{-0.04}^{+0.02}$\\
~~~Mass\dotfill & $M_{p}$  (\unit{M_{\oplus}}) \dotfill & $\cdots$ & $3\sigma < 99.2$\\
~~~Radius\dotfill & $R_{p}$  (\unit{R_{\oplus}}) \dotfill & $\cdots$ & $0.51 \pm 0.03$\\
~~~Semi-major Axis\dotfill & $a$ (au) \dotfill & $\cdots$ & $0.0080 \pm 0.0003$\\
~~~Average Incident Flux\dotfill & $\langle F \rangle$ (\unit{10^8\ erg/s/cm^2})\dotfill & $\cdots$ & $4.4 \pm 0.3$\\
~~~Equilibrium Temperature\(^{d}\)\dotfill & $T_{eq}$ (K)\dotfill & $\cdots$ & $1180 \pm 20$\\
\enddata
\tablenotetext{a}{$\mathcal{U}(a,b)$ is a uniform distribution defined between lower limit $a$ and upper limit $b$.}
\tablenotetext{b}{$\mathcal{J}(a,b)$ is a log-uniform distribution defined between lower limit $a$ and upper limit $b$.}
\tablenotetext{c}{$\mathcal{N}(a,b)$ is a normal distribution with mean $a$ and standard deviation.}
\tablenotetext{d}{The planet is assumed to be a black body.}
\end{deluxetable*}

\section{Discussion}\label{sec:disc}
\subsection{Constraints on Long-period Companions}
We use {\tt thejoker} \citep{Price-Whelan2017} to perform a rejection sampling analysis on the HPF RVs and place a constraint on the existence of long-period giant companions. \texttt{thejoker} is able to draw samples from the full posterior probability density function over orbital parameters and, despite the simplicity of the noise model (white noise) and single-companion assumption, it can be useful in characterizing the presence of massive planetary companions. \texttt{thejoker} uses a log-uniform prior for the period (between $0.4<P<410$ days), the Beta distribution (with $a = 0.867$ and $b = 3.03$) from \cite{Kipping2013a} as a prior for the eccentricity, and a uniform prior for the argument of pericenter and the orbital phase. For the rejection sampling analysis, we ran \(>\times10^8\) (\(2^{27}\)) samples with {\tt thejoker} exploring orbits with periods less than the RV baseline ($P<410$ days) . 

A total of 2194159 samples survived (\(\sim1.63\%\) acceptance rate). We calculate the masses for the surviving samples assuming $\sin i=1$ and place an upper limit (99th quantile) of $6.7\mathrm{M_{J}}$ for companions with periods $<410$ days ($<0.87$ au). If we only permit circular orbits, the corresponding upper limit is $4.7\mathrm{M_{J}}$. The HPF RVs suggest there are no low-inclination (\(\sin i=1\)) Jupiter-mass gas giants within 0.87 AU of KOI-4777. These constraints from HPF RVs are a useful to probe the existence of additional non-transiting companions orbiting KOI-4777 because no additional transits were detected by the Kepler DR25 pipeline or \cite{Caceres2019}.

\subsection{Bulk Composition of KOI-4777.01}
\begin{figure*}[!ht]
\epsscale{1.15}
\plotone{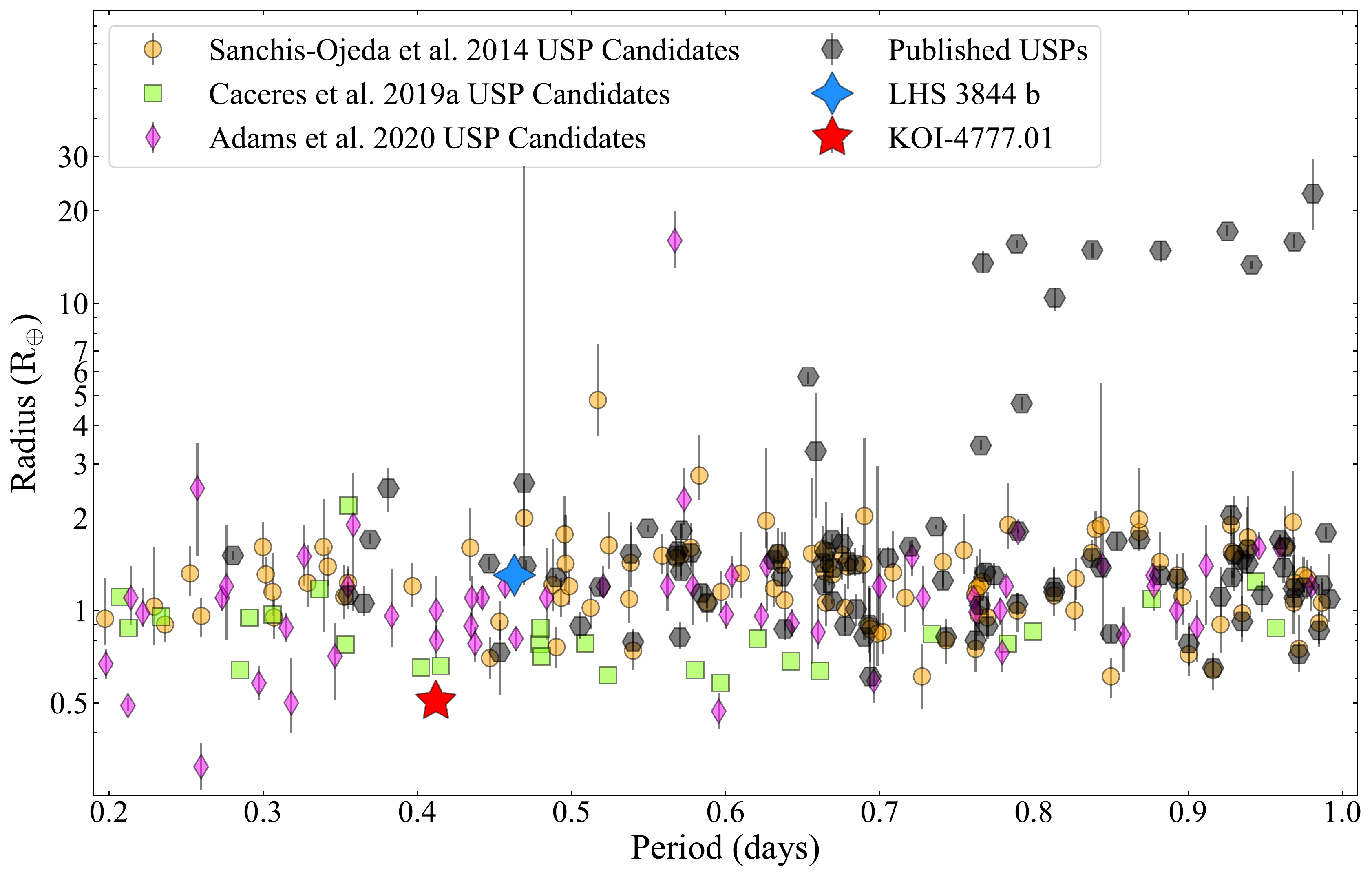}
\caption{KOI-4777.01 compared in orbital period and planetary radius with other candidate USPs from Kepler \citep{Sanchis-Ojeda2014,Caceres2019a} and K2 \citep{Adams2021}. USPs from literature were queried from the NASA Exoplanet Archive \citep[][]{Akeson2013} on 3 September 2021. LHS 3844 b \citep{Vanderspek2019} is marked as the blue star and another USP orbiting a similar M dwarf. KOI-4777.01, with a radius of $0.51 \pm 0.03~\mathrm{R_\oplus}$, is the smallest validated USP.}
\label{fig:usps}
\end{figure*}

KOI-4777.01 is on the extreme end of the USP population as it is the smallest validated planet in this population (Figure \ref{fig:usps}). Almost all validated and candidate USPs have radii $\lesssim 1.9 \mathrm{R}_{\oplus}$ and are expected to be largely rocky planets \citep[e.g.,][]{Sanchis-Ojeda2014,Adams2021}. \cite{Dai2019} performed a homogeneous study of USP Earth-sized planets and found that USPs with $M_{p}<8\mathrm{M}_{\oplus}$ were consistent with having Earth-like compositions. While our upper mass limit from RVs is much larger than $8M_\oplus$, a more informative bound on the mass can be obtained by placing constraints on the bulk density. The minimum density for a USP can be approximated by requiring it to orbit outside the Roche limit \citep[Equation 5 in][]{Rappaport2013} to ensure that it is not destroyed by the host star's tidal gravitational force. Following the assumption used by \cite{Sanchis-Ojeda2014}, in which the USP's core density is no more than twice its bulk density, the minimum density of KOI-4777.01 is $\rho_p\geq1.31~\mathrm{g/cm^3}$, which corresponds to a minimum mass of $M_p\geq0.025\mathrm{M_\oplus}$. To derive an upper density limit, the bulk density of KOI-4777 is required to be less than the density for the 100\% iron model from \cite{Zeng2019}. 

Using our upper $3\sigma$ radius measurement, we constrain KOI-4777.01 to $M_p\leq0.34\mathrm{M_\oplus}$. If instead, we force the upper density limit to be that of an Earth-like rocky planet \citep[32.5\% iron-nickel alloy and 67.5\% enstatite rock;][]{Zeng2019}, then KOI-4777.01 would have a mass $M_p\leq0.18\mathrm{M_\oplus}$. Both of these upper limits are larger than the mass of Mars ($0.1074~\mathrm{M_\oplus}$).

\begin{figure*}[!ht]
\epsscale{1.15}
\plotone{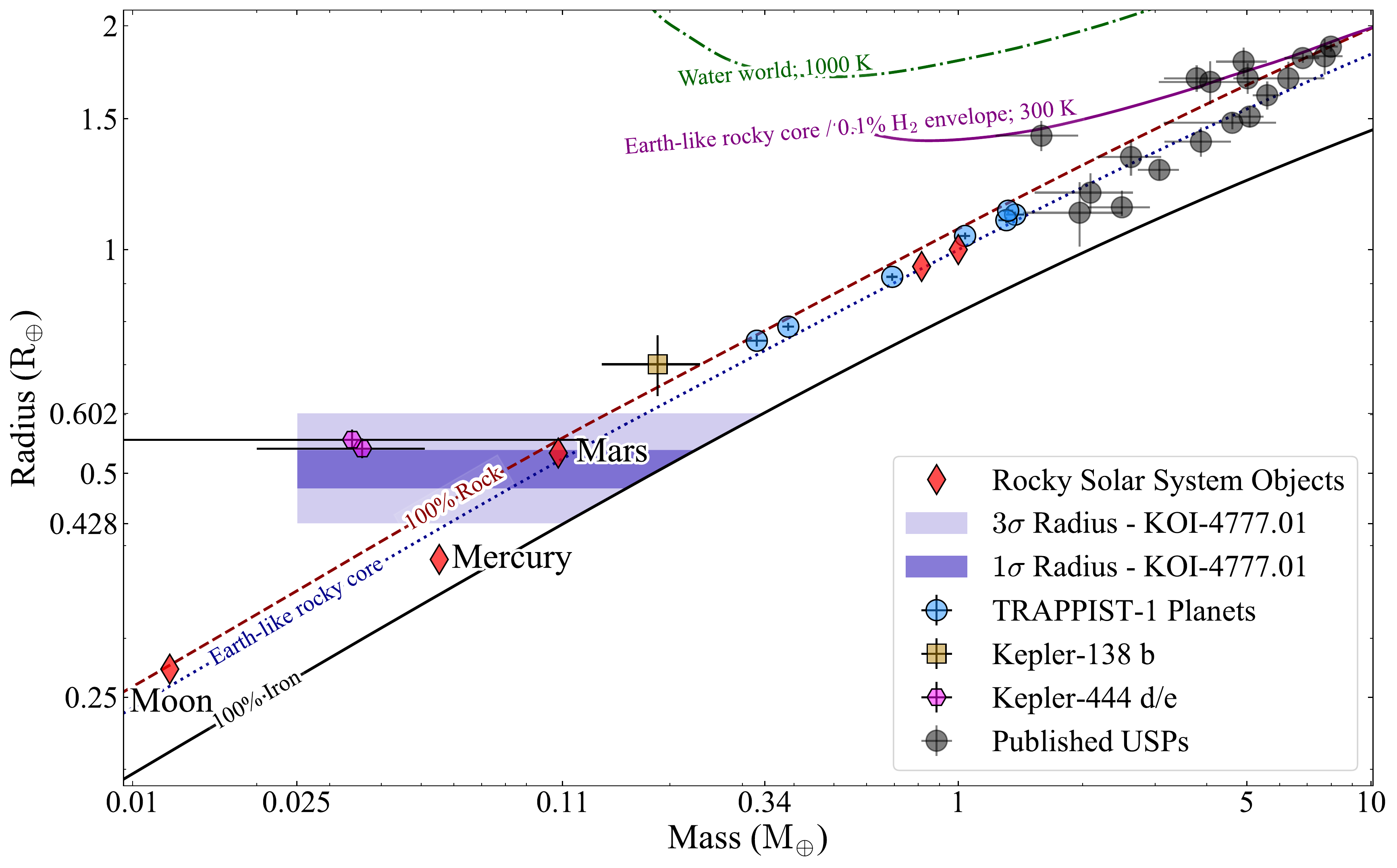}
\caption{KOI-4777.01 on the mass-radius diagram. Without a mass measurement, we only constrain the most probable location of KOI-4777.01. The shaded regions indicate our $1\sigma$ and $3\sigma$ radius measurement, and incorporate limits on the mass from density constraints. For comparison, we include the inner Solar System planets and the Moon (red diamonds), well-characterized and small Kepler planets \citep{Almenara2018,Mills2017}, the TRAPPIST-1 system \citep{Agol2021}, and known USPs (gray points) from the NASA Exoplanet Archive (queried 3 September 2021). We include several model composition curves from \cite{Zeng2019}. KOI-4777.01 is hot ($T_{eq}>1000$K), and is incompatible with models that include an atmosphere with temperature effects. It is compatible with models of rocky planets with no significant volatile envelope.}
\label{fig:mrad}
\end{figure*}

Given the mass regime of M$_{p}\leq0.34$M$_{\oplus}$, we expect KOI-4777.01 to be rocky and without a substantial volatile envelope. Figure \ref{fig:mrad} compares the calculated radius to known composition models to further investigate the bulk properties of KOI-4777.01. For comparison, we highlight published USPs contained in the NASA Exoplanet Archive and systems characterized with transit timing variations, such as the Kepler-444 Mars-sized planets \citep{Mills2017}, Kepler-138 b \citep{Almenara2018}, and the TRAPPIST-1 system \citep{Agol2021}. While the $1\sigma$ radius estimate of Kepler-138 b, Kepler-444 d, and Kepler-444 e all overlap with our $3\sigma$ radius measurement for KOI-4777.01, the aforementioned planets differ in the amount of insolation flux received. Kepler-138 b has a period of \(\sim 10.3\) days and is found to be consistent with the presence of a thick volatile layer while the Kepler-444 planets have periods \(>5\) days and could not exclude composition models containing volatiles. KOI-4777.01 is inconsistent with non-rocky compositions with a substantial hydrogen, helium, or water atmosphere. No model from \cite{Zeng2019} with an atmospheric temperature $1000$ K is within $3\sigma$ of our expected radius and this is consistent with simulations by \cite{Lopez2017} which show that KOI-4777.01 is too irradiated ($F>300F_\oplus$, $T_{eq}>1000$K) and too small to hold onto any H/He envelope. A pure rocky or Earth-like composition passes through the mass and radius envelope (shaded region in Figure \ref{fig:mrad}) for KOI-4777.01. The recent analysis of well-characterized USPs by \cite{Dai2021} showed that USPs tend to be rocky and cluster around an Earth-like composition with iron core mass fractions of \(0.32\pm0.04\). 

Planets acquire atmospheres either through direct nebular gas capture, planetesimal or cometary impacts, degassing during accretion, or volcanic outgassing \citep{Elkins-Tanton2008}. Low-mass planets, like KOI-4777.01, cannot effectively capture enough primordial gas to build large initial envelopes. For present-day Earth and Mars ingassing from the stellar nebula is not considered to be the primary contributor to the mass of their atmospheres \citep[e.g.,][]{Dauphas2013,Olson2019}. Additional factors would further limit the retention of a significant primary atmosphere, including stripping due to large impactors \citep{Schlichting2018} or hydrodynamic escape due to extreme UV and X-ray flux from a young stellar host \citep{Sharp2017}. Work by \citet{Kite2020} suggests that secondary atmospheres on rocky planets near an M dwarf host are generally unfavorable when $T_{e}>500$ K, in line with prior work describing the ``cosmic shoreline'' between planets with and without atmospheres \citep{Zahnle2017}. 

LHS 3844 b \citep{Vanderspek2019} is comparable to KOI-4777.01. While it is larger than KOI-4777.01 with a radius of $\sim$1.3 R$_{\oplus}$, LHS 3844 b is orbiting an M dwarf on a period of 11 hours and has a comparable equilibrium temperature of 805 K. \cite{Kreidberg2019} observed LHS 3844 b with Spitzer and obtained a symmetric thermal phase curve with a large amplitude that was inconsistent with the presence of a thick atmosphere. The Spitzer thermal phase curve could be modeled assuming LHS 3844 b was a synchronously-rotating bare rock with a surface composition comparable to Mercury (largely basaltic). \cite{Kane2020} modeled LHS 3844 b and characterized it as a bare rock planet and suggested a volatile-poor composition. We also expect the rotational period of KOI-4777.01 to be synchronous with the orbital frequency because the timescale for spin synchronization is shorter than the timescale for circularization even if we adopt an extreme value for KOI-4777.01's initial spin angular frequency \citep[\(\omega_{p}/n=10^4\) in Equation 5 of][]{Bodenheimer2001}. As such, KOI-4777.01 may also exhibit stark contrasts between the day- and night-side temperatures if it does not possess an appreciable atmosphere.

It is also possible that KOI-4777.01 may have an atmosphere made up of high molecular weight species. Following accretion, impact rates decline and smaller impactors can instead contribute to the planet's volatile inventory \citep[e.g.,][]{Genda2005, deVries2016}. Likewise, the extreme UV decreases over longer timescales as the host star settles down \citep[e.g.,][]{Sanz-Forcada2011}. Together with the relatively slow escape of heavier atoms \citep[e.g.,][]{Gronoff2020}, this can produce secondary atmospheres for terrestrial planets dominated by H$_{2}$O and CO$_{2}$, much like those hypothesized for early Venus and Earth \citep[e.g.,][]{Hamano2013}. Alternatively, a long-lived magma ocean or substellar pond would permit rock vapor into the atmosphere, potentially contributing detectable quantities of Na, Fe, SiO, and Mg to the atmosphere, depending on the surface temperature \citep[e.g.,][]{Schaefer2009, Costa2017}, although it is cooler than other objects with evidence for such an atmosphere \citep{Frustagli2020}. KOI-4777.01 receives a high enough insolation flux that there could be surface melting even in the absence of an atmosphere \citep{Chao2021}, especially if the planet experiences tidal or inductive heating driven by its host star. The escape of these species from a magma ocean-derived atmosphere could result in a trailing comet-like tail and an asymmetric transit shape \citep{Bodman2018}. Given the magnitude of the transit depth itself ($\sim$121 ppm), however, the spectroscopic characterization of the planet's atmosphere and inferences based on transit shape likely lies beyond the near-term capabilities of ground- and space-based observatories.

\subsection{Prospects for Future Characterization}
The faintness of the host star ($V=16.4$, $J=13.2$) and the small transit depth (\(\sim121\) ppm) make additional photometric characterization of KOI-4777 difficult. Photometric observations from the ground are currently impossible given this small transit depth, and the precision required to detect a transit is beyond the capabilities of more recent space-based missions, such as TESS. No occultations of KOI-4777.01 are observed in the Kepler data, and this is not surprising as the eclipse depth is $<1$ ppm in the Kepler bandpass. 

A mass determination of KOI-4777 is also difficult. If we adopt the upper mass limit of $0.34 \mathrm{M_\oplus}$, we expect a semi-amplitude velocity of $K=53$ cm/s which is beyond the reach of current instruments for such a faint target. The primary goal of additional RVs would be to further constrain the existence of long-period companions. KOI-4777 is an early M dwarf, and the RV information content for such stars \citep[e.g.,][]{Reiners2018} is better matched to optical spectrographs with an extended red wavelength coverage, such as CARMENES \cite{Quirrenbach2014,Quirrenbach2018}, MAROON-X \citep{Seifahrt2016}, or NEID \citep{Schwab2016}. Instruments with better precision can place tighter constraints on the existence of gas giants well beyond the ice-line for KOI-4777, and this is necessary to probe various formation scenarios. 

\section{Summary}\label{sec:summary}
We have validated the planetary nature of the smallest known USP, a Mars-sized exoplanet transiting KOI-4777, an early M dwarf that was recovered from false positive status via manual vetting by the Kepler FPWG and independently identified by the ARPS analysis. Despite the prevalence of USPs in multi-planet systems, HPF RVs do not reveal the presence of any low inclination (\(\sin i =1\)) Jupiter-mass gas giant within 0.87 AU of KOI-4777. More precise RV observations could tighten these constraints. While additional ground and space-based characterization of KOI-4777.01 is beyond the precision capabilities of current photometric and RV instruments, we place limits on the mass using density limits. The span of masses and radii for KOI-4777.01 suggest it is most likely depleted of any extensive atmosphere. 

\section*{acknowledgments}
This work was supported by NASA Headquarters under the NASA Earth and Space Science Fellowship Program through grant 80NSSC18K1114 and by the Alfred P. Sloan Foundation's Minority Ph.D. Program through grant G-2016-20166039. 
The Center for Exoplanets and Habitable Worlds is supported by the Pennsylvania State University and the Eberly College of Science.

This is University of Texas Center for Planetary Systems Habitability Contribution \#0038. These results are based on observations obtained with the Habitable-zone Planet Finder Spectrograph on the HET. We acknowledge support from NSF grants AST 1006676, AST 1126413, AST 1310875, AST 1310885, AST 2009889, AST 2108512 and the NASA Astrobiology Institute (NNA09DA76A) in our pursuit of precision radial velocities in the NIR. We acknowledge support from the Heising-Simons Foundation via grant 2017-0494.  The Hobby-Eberly Telescope is a joint project of the University of Texas at Austin, the Pennsylvania State University, Ludwig-Maximilians-Universität München, and Georg-August Universität Gottingen. The HET is named in honor of its principal benefactors, William P. Hobby and Robert E. Eberly. The HET collaboration acknowledges the support and resources from the Texas Advanced Computing Center. We are grateful to the HET Resident Astronomers and Telescope Operators for their valuable assistance in gathering our HPF data. 
We are grateful to the HET Resident Astronomers and Telescope Operators for their valuable assistance in gathering our HPF data. We would like to acknowledge that the HET is built on Indigenous land. Moreover, we would like to acknowledge and pay our respects to the Carrizo \& Comecrudo, Coahuiltecan, Caddo, Tonkawa, Comanche, Lipan Apache, Alabama-Coushatta, Kickapoo, Tigua Pueblo, and all the American Indian and Indigenous Peoples and communities who have been or have become a part of these lands and territories in Texas, here on Turtle Island.

Computations for this research were performed on the Pennsylvania State University's Institute for Computational and Data Sciences' Roar supercomputer, including the CyberLAMP cluster supported by NSF grant MRI-1626251.
The AutoRegressive Planet Search project at Penn State is supported by grants 80NSSC17K0122 from NASA and AST-1614690 from NSF.  

Some of the data presented in this paper were obtained from from the Mikulski Archive for Space Telescopes (MAST) at the Space Telescope Science Institute. The specific observations analyzed can be accessed via \dataset[10.17909/t9-nqgs-a902]{https://doi.org/10.17909/t9-nqgs-a902}. Support for MAST for non-HST data is provided by the NASA Office of Space Science via grant NNX09AF08G and by other grants and contracts. 
This work includes data collected by the Kepler mission, which are publicly available from MAST. Funding for the Kepler mission is provided by the NASA Science Mission directorate.

This research made use of the NASA Exoplanet Archive, which is operated by Caltech, under contract with NASA under the Exoplanet Exploration Program.
This research has made use of the SIMBAD database, operated at CDS, Strasbourg, France, and NASA's Astrophysics Data System Bibliographic Services.
2MASS is a joint project of the University of Massachusetts and IPAC at Caltech, funded by NASA and the NSF.

The Digitized Sky Surveys were produced at STSci under U.S. Government grant NAG W-2166. The images of these surveys are based on photographic data obtained using the Oschin Schmidt Telescope on Palomar Mountain and the UK Schmidt Telescope. The plates were processed into the present compressed digital form with the permission of these institutions. The National Geographic Society - Palomar Observatory Sky Atlas (POSS-I) was made by the California Institute of Technology with grants from the National Geographic Society. The Oschin Schmidt Telescope is operated by the California Institute of Technology and Palomar Observatory.

This work has made use of data from the European Space Agency (ESA) mission {\it Gaia} (\url{https://www.cosmos.esa.int/gaia}), processed by the {\it Gaia} Data Processing and Analysis Consortium (DPAC, \url{https://www.cosmos.esa.int/web/gaia/dpac/consortium}). Funding for the DPAC has been provided by national institutions, in particular the institutions participating in the {\it Gaia} Multilateral Agreement.

Some observations were obtained from the Pan-STARRS1 Surveys (PS1). PS1 and the PS1 public science archive have been made possible through contributions by the Institute for Astronomy, the University of Hawaii, the Pan-STARRS Project Office, the Max-Planck Society and its participating institutes, the Max Planck Institute for Astronomy, Heidelberg and the Max Planck Institute for Extraterrestrial Physics, Garching, The Johns Hopkins University, Durham University, the University of Edinburgh, the Queen's University Belfast, the Harvard-Smithsonian Center for Astrophysics, the Las Cumbres Observatory Global Telescope Network Incorporated, the National Central University of Taiwan, the Space Telescope Science Institute, the National Aeronautics and Space Administration under Grant No. NNX08AR22G issued through the Planetary Science Division of the NASA Science Mission Directorate, the National Science Foundation Grant No. AST-1238877, the University of Maryland, Eotvos Lorand University (ELTE), the Los Alamos National Laboratory, and the Gordon and Betty Moore Foundation.

\facilities{Gaia, HET (HPF), KPNO:2.1m (Robo-AO), PS1, Kepler} 
\software{
\texttt{ARC2} \citep{Aigrain2017},
\texttt{astroquery} \citep{Ginsburg2019},
\texttt{astropy} \citep{AstropyCollaboration2018},
\texttt{barycorrpy} \citep{Kanodia2018}, 
\texttt{batman} \citep{Kreidberg2015},
\texttt{dynesty} \citep{Speagle2020},
\texttt{EXOFASTv2} \citep{Eastman2019},
\texttt{galpy} \citep{Bovy2015},
\texttt{HPF-SpecMatch},
\texttt{juliet} \citep{Espinoza2019},
\texttt{matplotlib} \citep{hunter2007},
\texttt{numpy} \citep{vanderwalt2011},
\texttt{pandas} \citep{McKinney2010},
\texttt{radvel} \citep{Fulton2018},
\texttt{scipy} \citep{Virtanen2020},
\texttt{SERVAL},
\texttt{telfit} \citep{Gullikson2014},
\texttt{VESPA} \citep{Morton2012}
}

\bibliography{combined}

\end{document}